# RADIATIVE HEAT TRANSFER IN A PARALLELOGRAM SHAPED CAVITY

V. Le Dez and H. Sadat

Institut P', Université de Poitiers, Centre National de la Recherche Scientifique, Ecole Nationale Supérieure de Mécanique et Aérotechnique, 2 Rue Pierre Brousse, Bâtiment B25, TSA 41105, 86073 Poitiers Cedex 9, France

**Abstract:** An exact analytical description of the internal radiative field inside an emitting-absorbing gray semi-transparent medium enclosed in a two-dimensional parallelogram cavity is proposed. The expressions of the incident radiation and the radiative flux field are angularly and spatially discretized with a double Gauss quadrature, and the temperature field is obtained by using an iterative process. Some numerical solutions are tabulated and graphically presented as the benchmark solutions. Temperature and two components of the radiative flux are finally sketched on the whole domain. It is shown that the proposed method gives perfectly smooth results.

## Nomenclature

| | |
|---|---|
| $Bis_n$, $Cis_n$ | Altaç angular integrated Bickley-Naylor functions |
| $(\vec{e_x}, \vec{e_y}, \vec{e_z})$ | unit vectors of the *x*, *y*, *z* directions |
| $G$ | volumic incident radiation (Wm$^{-3}$) |
| $H_x$ | length of the cavity sides along the *x* direction (m) |
| $H_y$ | length of the cavity sides along the *y* direction (m) |
| $(i, j)$ | internal cells numbering |
| $I_{ij}(\vec{\Omega})$ | intensity at the $(i, j)$ cell centre (Wm$^{-2}$Sr$^{-1}$) |
| $Ki_n$ | Bickley-Naylor functions |
| $N_x$ | cells number on the sides parallel to the *x* direction |
| $N_y$ | cells number on the sides parallel to the *y* direction |
| $\vec{q}_{ij}^{\,r}$ | radiative flux vector at the $(i, j)$ cell centre (Wm$^{-2}$) |
| $q^x$ | *x*-component of the radiative flux (Wm$^{-2}$) |
| $q^y$ | *y*-component of the radiative flux (Wm$^{-2}$) |
| $S$ | volumic radiative source (Wm$^{-3}$) |
| $T$ | temperature (K) |
| *x, y, z* | coordinate axis directions |

*Greek letters*

| | |
|---|---|
| $\Delta x$ | characteristic cell length along the *x* direction (m) |
| $\Delta y$ | characteristic cell length along the *y* direction (m) |

| $\kappa$ | absorption coefficient (m$^{-1}$) |
| $\sigma$ | Stephan-Boltzmann constant (5.67 $10^{-8}$ Wm$^{-2}$K$^{-4}$) |
| $\tau$ | optical depth |
| $\varphi, \theta$ | angular description of the unit vector $\vec{\Omega}$ |
| $\vec{\Omega}$ | unit vector of radiation propagation |

*Subscripts (superscripts)*

E, N, O, S     east, north, west and south

## I – INTRODUCTION

Radiative effects are important in a large class of coupled thermal problems. This has led to the development of several numerical techniques to solve the radiative transfer equation in complex geometries. The particular case of the parallelogram shaped cavity has some interesting applications in buildings and solar energy systems [1]. Natural convection studies [2-4] have been conducted in such geometries and have shown the influence of the angle between adjacent boundaries on the flow pattern and heat transfer. Magnetic effects on the convection have also been studied and large modification of the flow structure has been observed [5]. The effect of radiative transfer due to emitting-reflecting surfaces has been considered by using the radiosity technique [6]. Baïri et al. [6] note that in such a cavity, presence of radiation incoming from the surfaces strongly affects the natural convection and may reduce it substantially for particular angles. To the best of our knowledge, radiative transfer when the medium in the cavity is participating has not been reported yet. The main goal of this paper is therefore to present accurate benchmark results which can be used to validate the results obtained by numerical methods. In the present paper we completely describe the radiative field inside a semi-transparent medium bounded by a parallelogram shaped cavity in a partially analytic way by keeping a hybrid formulation combining space and angular integrals as in [7]. The numerical treatment combines a discretization of the useful integrals and an iterative scheme to compute the temperature field at radiative equilibrium.

In the following, we first develop in section 2 the exact expressions of the radiative source and flux filed when using a hybrid formulation combining spatial and angular integrals. We then describe the angular and spatial discretizations of the useful integrals. Finally we present some numerical results in section 3 and end this work by a short conclusion.

## II – MATHEMATICAL FORMULATION

One considers an infinite parallelepiped with a parallelogram section, filled with an absorbing-emitting but non scattering semi-transparent gray medium, of absorption coefficient κ and unit refractive index at radiative equilibrium. The boundary surfaces are assumed isothermal with imposed temperatures and black for sake of simplicity, while the optical constants of the gray medium are not depending on the internal temperature field. The parallelogram section is divided into $N_x \times N_y$ isothermal parallelogram cells of equal lengths $\Delta x = \frac{H_x}{N_x}$ and $\Delta y = \frac{H_y}{N_y}$, where $H_x$ and $H_y$ are the two characteristic lengths of the parallelogram section, each of one labelled $(i, j)$, with $(i, j) \in \{1, ..., N_x\} \times \{1, ..., N_y\}$. The only considered energy transfer is radiation, whence the internal temperature field inside the parallelepiped is determined from the radiative equilibrium condition. The incident radiation and the radiative flux at a given internal point are given by:

$$G = \int_{\Omega=4\pi} I_{ij}(\vec{\Omega}) d\Omega$$
$$\vec{q_r} = \int_{\Omega=4\pi} I_{ij}(\vec{\Omega}) \vec{\Omega} d\Omega \qquad (1)$$

$I_{ij}(\vec{\Omega})$ is the radiative intensity at the centre $M_{ij}$ of the cell labelled $(i, j)$ for a given direction of propagation $\vec{\Omega}$. For semi-transparent gray media with a constant unit refractive index, the radiative source is simply the divergence of the radiative flux, expressed by:

$$S = \kappa(4\sigma T^4 - G) \qquad (2)$$

The temperature field at radiative equilibrium is then deduced from the incident radiation field.
Let us consider $(\vec{e_x}, \vec{e_y})$ the orthogonal basis of the parallelogram section, and $\vec{e_z}$ the unit vector orthogonal to $(\vec{e_x}, \vec{e_y})$. We note $\varphi$ the angle between the projection of a luminous ray on the parallelogram section and the unit vector $\vec{e_x}$, as illustrated in Fig. 1, while $\theta$ denotes the angle between the luminous ray and the unit vector $\vec{e_z}$ perpendicular to the figure's plane.
Since the parallelepiped is infinite in the $\vec{e_z}$ direction, the temperature field so as the radiative field do not depend on the z coordinate, and the contribution for angles $\theta \in \left[0, \frac{\pi}{2}\right]$ is strictly equivalent to the one for angles $\theta \in \left[\frac{\pi}{2}, \pi\right]$, whence the incident radiation writes in these conditions:

$$G = \int_{\theta=0}^{\pi} \int_{\varphi=0}^{2\pi} I_{ij}(\theta, \varphi) \sin\theta \, d\theta \, d\varphi = 2 \int_{\theta=0}^{\frac{\pi}{2}} \int_{\varphi=0}^{2\pi} I_{ij}(\theta, \varphi) \sin\theta \, d\theta \, d\varphi \qquad (3)$$

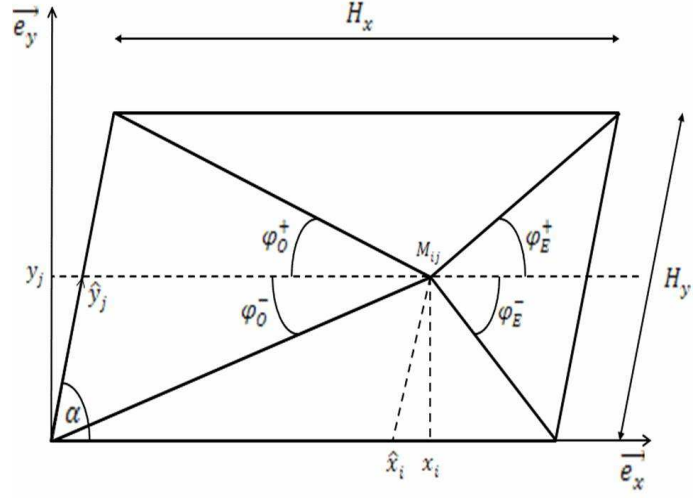

Figure 1: Geometry of the parallelogrammic cavity

We define four angular sectors in the plane $(\vec{e_x}, \vec{e_y})$ from the point $M_{ij}$ delimiting the radiative contributions originating from the cavity's surfaces. In Eq (3) the angle $\varphi$ is the natural angle defining the propagation direction $\vec{\Omega} = \begin{pmatrix} \cos\varphi \sin\theta \\ \sin\varphi \sin\theta \\ \cos\theta \end{pmatrix}$ in the natural basis $(\vec{e_x}, \vec{e_y}, \vec{e_z})$, passing through a given point $M_{ij}$. This means for example that given a particular angle $\varphi \in [0, \varphi_E^+]$ where the geometrical angular sector aperture $\varphi_E^+$ is represented in Fig. 1, the radiation incoming at point $M_{ij}$ for this direction is originating from the southern surface and reaches the eastern surface, and is not coming from the eastern one. Similarly, given an angle $\varphi \in [2\pi - \varphi_E^-, 2\pi] \equiv [-\varphi_E^-, 0]$, where the absolute value angular aperture $\varphi_E^-$ is shown in Fig. 1, the incoming radiation is originating from the northern surface. We shall therefore use a ray tracing process to calculate both the incident radiation and radiative flux, by considering the opposite directions $-\vec{\Omega}$: doing so is equivalent to consider that for a given angle $\varphi \in [0, \varphi_E^+]$, the effective radiation passing through $M_{ij}$ is incoming from the northern part of the eastern boundary and reaches then the southern surface. This calculation process, which is much simpler, has no incidence on the incident radiation $G$, but leads to the opposite radiative flux. This means that the determination of the incident radiation and radiative flux, is performed with the direction $-\vec{\Omega}$, and that the sign of the correct radiative flux is the opposite of the calculated one.

In what follows, we shall develop the determination of the eastern part of the incident radiation and radiative flux before giving a complete generalisation to the three other surfaces.

For an internal point $M$ such that $H_y \cos\alpha \leq x \leq H_x$, where $x$ is the natural abscissa along the axis $\vec{e_x}$, the eastern part of the incident radiation writes:

$$G_E = 2 \int_{\theta=0}^{\frac{\pi}{2}} \int_{\varphi=2\pi-\varphi_E^-}^{\varphi_E^+} I_{ij}(\theta, \varphi) \sin\theta \, d\theta \, d\varphi \tag{4}$$

Let us introduce the modified coordinates $\hat{x}$ and $\hat{y}$ along lines parallel to the eastern and western surfaces of the cavity, such that:

$$x = \hat{x} + \hat{y} \cos \alpha \qquad (5)$$
$$y = \hat{y} \sin \alpha$$

The two boundary angles are given by $\varphi_E^+ = \tan^{-1}\left[\frac{(H_y - \hat{y}) \sin \alpha}{H_x - \hat{x} + (H_y - \hat{y}) \cos \alpha}\right]$ and $\varphi_E^- = \tan^{-1}\left(\frac{\hat{y} \sin \alpha}{H_x - \hat{x} - \hat{y} \cos \alpha}\right)$. For a given direction $\vec{\Omega}$ inside this angular sector, the solution of the radiative transfer equation (RTE) between $M_{ij}$ and a corresponding point located on the eastern surface is given by:

$$I_{ij}(\theta, \varphi) = \frac{\sigma T_E^4}{\pi} e^{-\kappa \Delta} + \frac{\kappa \sigma}{\pi} \int_{s=0}^{\Delta} T^4(s) e^{-\kappa s} ds \qquad (6)$$

where $\Delta$ is the length between $M_{ij}$ and the intersection of the eastern surface and the line supported by $\vec{\Omega}$, as illustrated in Fig. 2

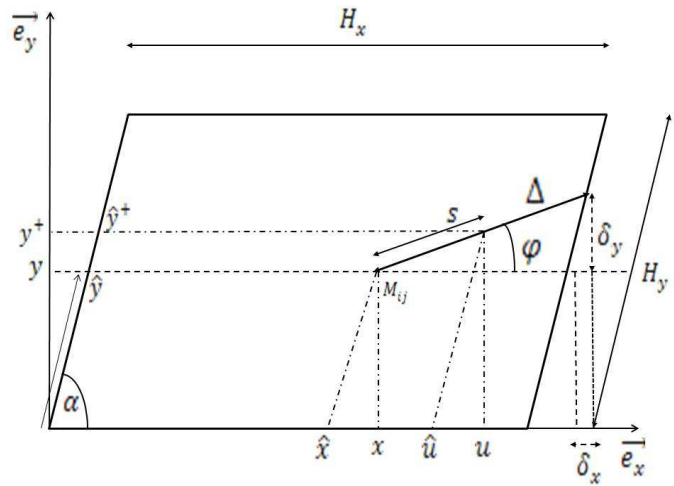

Figure 2: Determination of the geometrical elements for radiation incoming from the northern part of the eastern boundary

With the notations of Fig. 2, the useful geometrical quantities are $\cos \varphi = \frac{H_x - \hat{x} + \delta_x}{\Delta}$, $\sin \varphi = \frac{\delta_y}{\Delta}$ and $\tan \alpha = \frac{\delta_y}{\delta_x}$, whence $\delta_x = \frac{\Delta \sin \varphi}{\tan \alpha}$ and $\cos \varphi = \frac{H_x - \hat{x} + \frac{\Delta \sin \varphi}{\tan \alpha}}{\Delta}$. A simple elimination leads to the travel length $\Delta = \frac{(H_x - \hat{x}) \sin \alpha}{\sin(\alpha - \varphi)}$. This length evaluated in the plane of the figure, can directly be found by observing that $\Delta$ is the side of a triangle whose opposite angle is $\pi - \alpha$, while $H_x - \hat{x}$ is another side of the same triangle, whose opposite angle is $\alpha - \varphi$. Similarly, for a given parameter $\hat{u} \in [\hat{x}, H_x]$, the length $s$ along the line $\Delta$ between $\hat{x}$ and $\hat{u}$ is expressed by $s = \frac{(\hat{u}-\hat{x})\sin\alpha}{\sin(\alpha-\varphi)}$.

For a corresponding point $u$, the useful coordinate is: $y^+ = y + s\sin\varphi = \left[\hat{y} + \frac{(\hat{u}-\hat{x})\sin\varphi}{\sin(\alpha-\varphi)}\right]\sin\alpha$, whence $\hat{y}^+ = \hat{y} + \frac{(\hat{u}-\hat{x})\sin\varphi}{\sin(\alpha-\varphi)}$.

All these geometrical quantities have been obtained in the plane of the figure, for $\theta = \frac{\pi}{2}$. The parallelepiped of parallelogram section being infinite in the $\vec{e_z}$ direction, all the previous useful lengths have to be divided by $\sin\theta$ for a direction $\theta \neq \frac{\pi}{2}$, except for the ones used in the temperature field which is two-dimensional and not depending on the $z$ (and therefore $\theta$) coordinate.

In a discrete formulation, each parallelogram cell is supposed isothermal, which induces that the effective temperature field depends on the coordinates $\hat{x}$ and $\hat{y}$. The incoming intensity $I_{ij}(\theta, \varphi)$ for all space directions whose projection along $\vec{e_z}$ on the plane $(\vec{e_x}, \vec{e_y})$ is the line $\Delta$ is obtained from Eq. (6):

$$I_{ij}(\theta, \varphi) = \frac{\sigma T_E^4}{\pi} e^{-\frac{\kappa(H_x - \hat{x})\sin\alpha}{\sin(\alpha-\varphi)\sin\theta}} \\ + \frac{\kappa\sigma\sin\alpha}{\pi\sin(\alpha-\varphi)\sin\theta} \int_{\hat{u}=\hat{x}}^{H_x} T^4\left[\hat{u}, \hat{y} + \frac{(\hat{u}-\hat{x})\sin\varphi}{\sin(\alpha-\varphi)}\right] e^{-\frac{\kappa(\hat{u}-\hat{x})\sin\alpha}{\sin(\alpha-\varphi)\sin\theta}} d\hat{u} \quad (7)$$

Introducing the Bickley-Naylor functions [8] defined by:

$$Ki_n(z) = \int_{\theta=0}^{\frac{\pi}{2}} e^{-\frac{z}{\sin\theta}}(\sin\theta)^{n-1}\, d\theta \quad (8)$$

implies for the incident radiation and the radiative flux (with the correct sign) incoming from the north location of the eastern surface:

$$\frac{G_E^+}{2} = \frac{\sigma T_E^4}{\pi} \int_{\varphi=0}^{\varphi_E^+} Ki_2\left[\frac{\kappa(H_x - \hat{x})\sin\alpha}{\sin(\alpha-\varphi)}\right] d\varphi \\ + \frac{\kappa\sigma\sin\alpha}{\pi} \int_{\varphi=0}^{\varphi_E^+}\int_{\hat{u}=\hat{x}}^{H_x} T^4\left[\hat{u}, \hat{y} + \frac{(\hat{u}-\hat{x})\sin\varphi}{\sin(\alpha-\varphi)}\right] Ki_1\left[\frac{\kappa(\hat{u}-\hat{x})\sin\alpha}{\sin(\alpha-\varphi)}\right] d\hat{u}\, \frac{d\varphi}{\sin(\alpha-\varphi)} \quad (9)$$

$$-\frac{\vec{q}_E^+}{2} = \frac{\sigma T_E^4}{\pi} \int_{\varphi=0}^{\varphi_E^+} Ki_3 \left[\frac{\kappa(H_x - \hat{x}) \sin \alpha}{\sin(\alpha - \varphi)}\right] \begin{pmatrix} \cos \varphi \\ \sin \varphi \end{pmatrix} d\varphi$$
$$+ \frac{\kappa \sigma \sin \alpha}{\pi} \int_{\varphi=0}^{\varphi_E^+} \int_{\hat{u}=\hat{x}}^{H_x} T^4 \left[\hat{u}, \hat{y} + \frac{(\hat{u} - \hat{x}) \sin \varphi}{\sin(\alpha - \varphi)}\right] Ki_2 \left[\frac{\kappa(\hat{u} - \hat{x}) \sin \alpha}{\sin(\alpha - \varphi)}\right] d\hat{u} \begin{pmatrix} \cos \varphi \\ \sin \varphi \end{pmatrix} \frac{d\varphi}{\sin(\alpha - \varphi)} \quad (10)$$

where the two components $q_E^x$ and $q_E^y$ of the radiative vector are determined along the natural axes $\vec{e_x}$ and $\vec{e_y}$.

A similar calculation is performed to obtain the radiative contribution incoming from the southern part of the eastern surface, the corresponding situation being depicted in Fig. 3. The incident radiation and the radiative flux are here defined by:

$$G_E^- = 2 \int_{\theta=0}^{\frac{\pi}{2}} \int_{\varphi=2\pi-\varphi_E^-}^{2\pi} I_{ij}(\theta, \varphi) \sin \theta \, d\theta \, d\varphi = 2 \int_{\theta=0}^{\frac{\pi}{2}} \int_{\varphi=0}^{\varphi_E^-} I_{ij}(\theta, \varphi) \sin \theta \, d\theta \, d\varphi$$

$$-\vec{q}_E^- = 2 \int_{\theta=0}^{\frac{\pi}{2}} \int_{\varphi=2\pi-\varphi_E^-}^{2\pi} I_{ij}(\theta, \varphi) \sin \theta \, d\theta \begin{pmatrix} \cos \varphi \\ \sin \varphi \end{pmatrix} d\varphi = 2 \int_{\theta=0}^{\frac{\pi}{2}} \int_{\varphi=0}^{\varphi_E^-} I_{ij}(\theta, \varphi) \sin \theta \, d\theta \begin{pmatrix} \cos \varphi \\ -\sin \varphi \end{pmatrix} d\varphi$$

The distance between point $M_{ij}$ and the intersection point on the eastern surface is $\Delta = \frac{(H_x - \hat{x}) \sin \alpha}{\sin(\alpha + \varphi)}$, and for a parameter $\hat{u} \in [\hat{x}, H_x]$, the corresponding distance is $s = \frac{(\hat{u} - \hat{x}) \sin \alpha}{\sin(\alpha + \varphi)}$, while the effective ordinate is $\hat{y}^- = \hat{y} - \frac{(\hat{u} - \hat{x}) \sin \varphi}{\sin(\alpha + \varphi)}$.

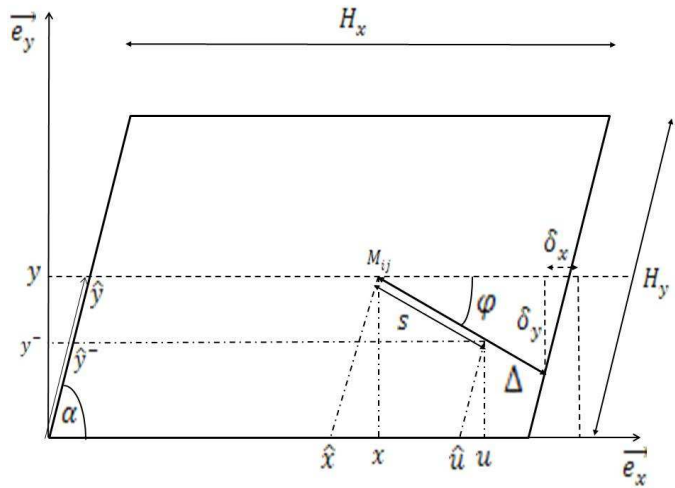

Figure 3: Determination of the geometrical elements for radiation incoming from the southern part of the eastern boundary

One deduces the southern part of the eastern surface contribution to the incident radiation and radiative vector components:

$$\frac{G_E^-}{2} = \frac{\sigma T_E^4}{\pi} \int_{\varphi=0}^{\varphi_E^-} Ki_2\left[\frac{\kappa(H_x - \hat{x}) \sin \alpha}{\sin(\alpha + \varphi)}\right] d\varphi$$
$$+ \frac{\kappa \sigma \sin \alpha}{\pi} \int_{\varphi=0}^{\varphi_E^-} \int_{\hat{u}=\hat{x}}^{H_x} T^4\left[\hat{u}, \hat{y} - \frac{(\hat{u} - \hat{x}) \sin \varphi}{\sin(\alpha + \varphi)}\right] Ki_1\left[\frac{\kappa(\hat{u} - \hat{x}) \sin \alpha}{\sin(\alpha + \varphi)}\right] d\hat{u} \frac{d\varphi}{\sin(\alpha + \varphi)}$$
(11)

$$-\frac{\vec{q}_E^-}{2} = \frac{\sigma T_E^4}{\pi} \int_{\varphi=0}^{\varphi_E^-} Ki_3\left[\frac{\kappa(H_x - \hat{x}) \sin \alpha}{\sin(\alpha + \varphi)}\right] \begin{pmatrix} \cos \varphi \\ \sin \varphi \end{pmatrix} d\varphi$$
$$+ \frac{\kappa \sigma \sin \alpha}{\pi} \int_{\varphi=0}^{\varphi_E^-} \int_{\hat{u}=\hat{x}}^{H_x} T^4\left[\hat{u}, \hat{y} - \frac{(\hat{u} - \hat{x}) \sin \varphi}{\sin(\alpha + \varphi)}\right] Ki_2\left[\frac{\kappa(\hat{u} - \hat{x}) \sin \alpha}{\sin(\alpha + \varphi)}\right] d\hat{u} \begin{pmatrix} \cos \varphi \\ -\sin \varphi \end{pmatrix} \frac{d\varphi}{\sin(\alpha + \varphi)}$$
(12)

The calculation of the different boundary terms is shown in the annex 1.

At this stage, one must take into account points $M_{ij}$ for which the abscissa $x$ is greater than $H_x$, such that $H_x < x \leq H_x + H_y \cos \alpha$. For such points, the angle $\varphi_E^-$ is greater than $\frac{\pi}{2}$, and the corresponding incident radiation and radiative flux are given by:

$$G_E^- = 2 \int_{\theta=0}^{\frac{\pi}{2}} \int_{\varphi=0}^{\pi - \tan^{-1}\left(\frac{\hat{y} \sin \alpha}{\hat{y} \cos \alpha - H_x + \hat{x}}\right)} I_{ij}(\theta, \varphi) \sin \theta \, d\theta \, d\varphi$$

$$\vec{q}_E^- = -2 \int_{\theta=0}^{\frac{\pi}{2}} \int_{\varphi=0}^{\pi - \tan^{-1}\left(\frac{\hat{y} \sin \alpha}{\hat{y} \cos \alpha - H_x + \hat{x}}\right)} I_{ij}(\theta, \varphi) \sin \theta \, d\theta \begin{pmatrix} \cos \varphi \\ -\sin \varphi \end{pmatrix} d\varphi$$

It is easy to verify that the calculation procedure used in the case $\varphi_E^- < \frac{\pi}{2}$ remains valid for determining the distances, whence the global incident radiation incoming from the whole eastern surface writes:

$$\frac{G_E}{2} = \frac{\sigma C_E T_E^4}{\pi} + \frac{\kappa \sigma \sin \alpha}{\pi} \left\{ \int_{\varphi=0}^{\varphi_E^+} \int_{\hat{u}=\hat{x}}^{H_x} T^4\left[\hat{u}, \hat{y} + \frac{(\hat{u} - \hat{x}) \sin \varphi}{\sin(\alpha - \varphi)}\right] Ki_1\left[\frac{\kappa(\hat{u} - \hat{x}) \sin \alpha}{\sin(\alpha - \varphi)}\right] d\hat{u} \frac{d\varphi}{\sin(\alpha - \varphi)} \right.$$
$$\left. + \int_{\varphi=0}^{\aleph_E} \int_{\hat{u}=\hat{x}}^{H_x} T^4\left[\hat{u}, \hat{y} - \frac{(\hat{u} - \hat{x}) \sin \varphi}{\sin(\alpha + \varphi)}\right] Ki_1\left[\frac{\kappa(\hat{u} - \hat{x}) \sin \alpha}{\sin(\alpha + \varphi)}\right] d\hat{u} \frac{d\varphi}{\sin(\alpha + \varphi)} \right\}$$
(13)

with:

$$\aleph_E = \tan^{-1}\left(\frac{\hat{y} \sin \alpha}{H_x - \hat{x} - \hat{y} \cos \alpha}\right) \quad \text{if} \quad H_x - \hat{x} \geq \hat{y} \cos \alpha$$
$$\aleph_E = \pi - \tan^{-1}\left(\frac{\hat{y} \sin \alpha}{\hat{y} \cos \alpha - H_x + \hat{x}}\right) \quad \text{if} \quad H_x - \hat{x} < \hat{y} \cos \alpha$$

and:

$$C_E = Bis_2[\kappa(H_x - \hat{x})\sin\alpha, \psi_E^+] + \delta_E Bis_2[\kappa(H_x - \hat{x})\sin\alpha, \omega_E]$$

Similarly the two components of the radiative flux vector finally write:

$$-\frac{q_E^x}{2} = \frac{\sigma D_E T_E^4}{\pi}$$
$$+ \frac{\kappa\sigma \sin\alpha}{\pi}\left\{\int_{\varphi=0}^{\varphi_E^+}\int_{\hat{u}=\hat{x}}^{H_x} T^4\left[\hat{u}, \hat{y} + \frac{(\hat{u}-\hat{x})\sin\varphi}{\sin(\alpha-\varphi)}\right] Ki_2\left[\frac{\kappa(\hat{u}-\hat{x})\sin\alpha}{\sin(\alpha-\varphi)}\right] d\hat{u}\frac{\cos\varphi\, d\varphi}{\sin(\alpha-\varphi)} \right.$$
$$\left. + \int_{\varphi=0}^{\aleph_E}\int_{\hat{u}=\hat{x}}^{H_x} T^4\left[\hat{u}, \hat{y} - \frac{(\hat{u}-\hat{x})\sin\varphi}{\sin(\alpha+\varphi)}\right] Ki_2\left[\frac{\kappa(\hat{u}-\hat{x})\sin\alpha}{\sin(\alpha+\varphi)}\right] d\hat{u}\frac{\cos\varphi\, d\varphi}{\sin(\alpha+\varphi)}\right\}$$

(14)

$$-\frac{q_E^y}{2} = \frac{\sigma E_E T_E^4}{\pi}$$
$$+ \frac{\kappa\sigma \sin\alpha}{\pi}\left\{\int_{\varphi=0}^{\varphi_E^+}\int_{\hat{u}=\hat{x}}^{H_x} T^4\left[\hat{u}, \hat{y} + \frac{(\hat{u}-\hat{x})\sin\varphi}{\sin(\alpha-\varphi)}\right] Ki_2\left[\frac{\kappa(\hat{u}-\hat{x})\sin\alpha}{\sin(\alpha-\varphi)}\right] d\hat{u}\frac{\sin\varphi\, d\varphi}{\sin(\alpha-\varphi)} \right.$$
$$\left. - \int_{\varphi=0}^{\aleph_E}\int_{\hat{u}=\hat{x}}^{H_x} T^4\left[\hat{u}, \hat{y} - \frac{(\hat{u}-\hat{x})\sin\varphi}{\sin(\alpha+\varphi)}\right] Ki_2\left[\frac{\kappa(\hat{u}-\hat{x})\sin\alpha}{\sin(\alpha+\varphi)}\right] d\hat{u}\frac{\sin\varphi\, d\varphi}{\sin(\alpha+\varphi)}\right\}$$

(15)

where:

$$D_E = \sin\alpha\,\{Bis_3[\kappa(H_x - \hat{x})\sin\alpha, \psi_E^+] + \delta_E Bis_3[\kappa(H_x - \hat{x})\sin\alpha, \omega_E]\}$$
$$+ \cos\alpha\,\{Cis_3[\kappa(H_x - \hat{x})\sin\alpha, \psi_E^+] - Cis_3[\kappa(H_x - \hat{x})\sin\alpha, \omega_E]\}$$

$$E_E = \sin\alpha\,\{Cis_3[\kappa(H_x - \hat{x})\sin\alpha, \psi_E^+] - Cis_3[\kappa(H_x - \hat{x})\sin\alpha, \omega_E]\}$$
$$- \cos\alpha\,\{Bis_3[\kappa(H_x - \hat{x})\sin\alpha, \psi_E^+] + \delta_E Bis_3[\kappa(H_x - \hat{x})\sin\alpha, \omega_E]\}$$

In the previous expressions, the $Bis_n$ and $Cis_n$ functions [8] are defined as follows:

$$Bis_n(x, \theta) = \int_{\varphi=0}^{\theta} Ki_n\left(\frac{x}{\cos\varphi}\right)(\cos\varphi)^{n-2} d\varphi \text{ and } Cis_n(x, \theta) = \int_{\varphi=0}^{\theta} Ki_n\left(\frac{x}{\cos\varphi}\right)(\cos\varphi)^{n-3}\sin\varphi\, d\varphi$$

The contributions from the three other surfaces which are not given here for conciseness, are determined in a very similar way. The incident radiation (and radiative flux) at a given point of the medium is the sum of contributions from each surface of the cavity.

The integrals appearing in the incident radiation expression must be calculated numerically. A powerful way is to express each integral as a linear combination of temperatures power four with a set of geometrical coefficients as in [7], but their determination is a difficult task for this kind of geometry. A simpler scheme can be obtained by transforming the angular and spatial integrals into a double sum with the help of a numerical quadrature as described in the annex 2.

## III – NUMERICAL RESULTS

We present in this section the results obtained for $H_x = H_y = 1m$. All but one black boundary temperatures are set to zero, the non-zero temperature being set to one on the southern side of the cavity, as in [9]. A Gauss quadrature is used to calculate both the angular and spatial integrals. A precise examination of the number of angular directions shows that for $N_\varphi \geq 21$ the results do no longer vary, whatever the absorption coefficient and number of cells are. Due to the repartition in four angular sectors (north, east, west and south) at each point, each sector being also divided in two distinct parts (for example the northern and southern propagation from the eastern surface), the numerical calculation takes into account 8 angular sectors at each point of the cavity, the total aperture of each angular sector being lower than 45°. Then the mean angle between each propagation direction is around 2° when computing all the angular integrals with a 21 points Gauss quadrature. The two cases of radiative equilibrium and of a given temperature field will be considered.

### III.1 Radiative equilibrium

The procedure performed to calculate the temperature field inside the cavity at radiative equilibrium consists in evaluating numerically at a first stage for a given temperature field, all the integrals appearing in the exact four expressions of the incident radiation in each discrete point inside the medium, from each surface. Then the temperature at each point is obtained from the radiative equilibrium condition: $T_{ij}^4 = \frac{G_{ij}}{4\sigma}$. Once the new temperature field is obtained, all the integrals of the incident radiation are re-evaluated with this field, and the process is achieved at convergence of the temperature field, which allows the computation of the radiative flux field.

To assess the validity of the proposed method, we have first determined the temperature field inside a square cavity [7, 9]. The results obtained which are not presented here for conciseness are accurate and similar to those reported in [7]. A uniform 151*151 grid was used for all the simulations, which insures smooth results for all examined cases.

We give in Tables 1a and 1b discrete values of the temperature power four on the medium line of the cavity $\hat{x} = \frac{1}{2}$ at regularly spaced ordinates positions $\hat{y}$ for six values of the edge angle $\alpha$ and for two absorption coefficients $\kappa = 1.\,m^{-1}$ and $\kappa = 5.\,m^{-1}$.

| $\hat{y}$ | $\alpha = 15°$ | $\alpha = 30°$ | $\alpha = 45°$ | $\alpha = 60°$ | $\alpha = 75°$ | $\alpha = 90°$ |
|---|---|---|---|---|---|---|
| 0.05 | 0.5498 | 0.5629 | 0.5673 | 0.5685 | 0.5687 | 0.5686 |
| 0.10 | 0.5324 | 0.5332 | 0.5286 | 0.5236 | 0.5202 | 0.5190 |
| 0.15 | 0.5152 | 0.5042 | 0.4917 | 0.4820 | 0.4761 | 0.4741 |
| 0.20 | 0.4967 | 0.4743 | 0.4554 | 0.4425 | 0.4352 | 0.4329 |
| 0.25 | 0.4759 | 0.4424 | 0.4191 | 0.4049 | 0.3974 | 0.3951 |
| 0.30 | 0.4506 | 0.4078 | 0.3830 | 0.3694 | 0.3626 | 0.3605 |
| 0.35 | 0.4182 | 0.3701 | 0.3473 | 0.3360 | 0.3305 | 0.3289 |
| 0.40 | 0.3746 | 0.3300 | 0.3128 | 0.3049 | 0.3012 | 0.3001 |
| 0.45 | 0.3168 | 0.2892 | 0.2801 | 0.2762 | 0.2744 | 0.2739 |
| 0.50 | 0.2500 | 0.2500 | 0.2500 | 0.2500 | 0.2500 | 0.2500 |
| 0.55 | 0.1891 | 0.2144 | 0.2225 | 0.2261 | 0.2277 | 0.2282 |
| 0.60 | 0.1432 | 0.1834 | 0.1978 | 0.2043 | 0.2073 | 0.2082 |
| 0.65 | 0.1109 | 0.1570 | 0.1757 | 0.1845 | 0.1886 | 0.1898 |
| 0.70 | 0.0879 | 0.1347 | 0.1560 | 0.1664 | 0.1713 | 0.1728 |
| 0.75 | 0.0712 | 0.1159 | 0.1384 | 0.1498 | 0.1553 | 0.1570 |
| 0.80 | 0.0585 | 0.0999 | 0.1226 | 0.1346 | 0.1404 | 0.1422 |
| 0.85 | 0.0487 | 0.0862 | 0.1082 | 0.1203 | 0.1264 | 0.1282 |
| 0.90 | 0.0409 | 0.0743 | 0.0951 | 0.1069 | 0.1129 | 0.1147 |
| 0.95 | 0.0345 | 0.0639 | 0.0829 | 0.0939 | 0.0996 | 0.1013 |

Table 1a: Temperature power 4 on the line $\hat{x} = \frac{1}{2}$ for different angles $\alpha$ at $\kappa = 1.\,m^{-1}$

| $\hat{y}$ | $\alpha = 15°$ | $\alpha = 30°$ | $\alpha = 45°$ | $\alpha = 60°$ | $\alpha = 75°$ | $\alpha = 90°$ |
|---|---|---|---|---|---|---|
| 0.05 | 0.6636 | 0.7034 | 0.7187 | 0.7249 | 0.7273 | 0.7278 |
| 0.10 | 0.6300 | 0.6526 | 0.6562 | 0.6547 | 0.6527 | 0.6519 |
| 0.15 | 0.5991 | 0.6052 | 0.5984 | 0.5907 | 0.5855 | 0.5837 |
| 0.20 | 0.5682 | 0.5582 | 0.5427 | 0.5307 | 0.5236 | 0.5212 |
| 0.25 | 0.5355 | 0.5102 | 0.4885 | 0.4744 | 0.4666 | 0.4642 |
| 0.30 | 0.4986 | 0.4602 | 0.4357 | 0.4217 | 0.4145 | 0.4123 |
| 0.35 | 0.4543 | 0.4079 | 0.3848 | 0.3729 | 0.3671 | 0.3653 |
| 0.40 | 0.3985 | 0.3540 | 0.3363 | 0.3279 | 0.3239 | 0.3228 |
| 0.45 | 0.3283 | 0.3005 | 0.2912 | 0.2870 | 0.2850 | 0.2844 |
| 0.50 | 0.2500 | 0.2500 | 0.2500 | 0.2500 | 0.2500 | 0.2500 |
| 0.55 | 0.1791 | 0.2045 | 0.2129 | 0.2167 | 0.2184 | 0.2189 |
| 0.60 | 0.1255 | 0.1652 | 0.1799 | 0.1867 | 0.1899 | 0.1909 |
| 0.65 | 0.0876 | 0.1319 | 0.1508 | 0.1599 | 0.1642 | 0.1655 |
| 0.70 | 0.0613 | 0.1043 | 0.1252 | 0.1358 | 0.1409 | 0.1425 |
| 0.75 | 0.0429 | 0.0815 | 0.1028 | 0.1141 | 0.1197 | 0.1215 |
| 0.80 | 0.0299 | 0.0628 | 0.0831 | 0.0945 | 0.1003 | 0.1021 |
| 0.85 | 0.0207 | 0.0474 | 0.0658 | 0.0767 | 0.0823 | 0.0841 |
| 0.90 | 0.0141 | 0.0349 | 0.0506 | 0.0603 | 0.0655 | 0.0671 |
| 0.95 | 0.0095 | 0.0247 | 0.0369 | 0.0448 | 0.0491 | 0.0505 |

Table 1b: Temperature power 4 on the line $\hat{x} = \frac{1}{2}$ for different angles $\alpha$ at $\kappa = 5.\,m^{-1}$

One can see that the exact value of 0.25 is obtained at the centre of the cavity for all angles. The results in temperatures for the square cavity have been compared to [7, 9] for different absorption coefficients and show an excellent agreement.

We present now in figure 4 the temperatures profiles for different angles α and on the three particular lines $\hat{x} = \frac{1}{4}$, $\hat{x} = \frac{1}{2}$ and $\hat{x} = \frac{3}{4}$ for the two absorption coefficients.

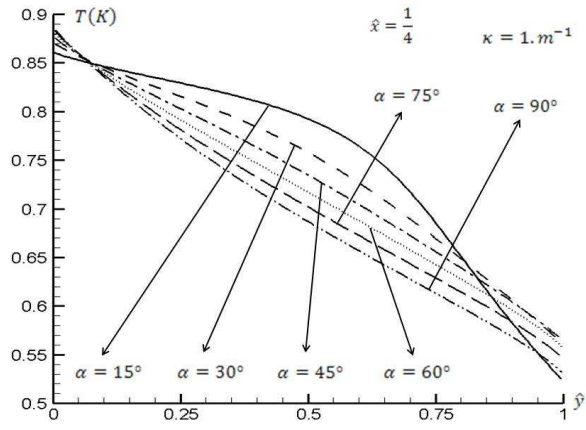
(a1)

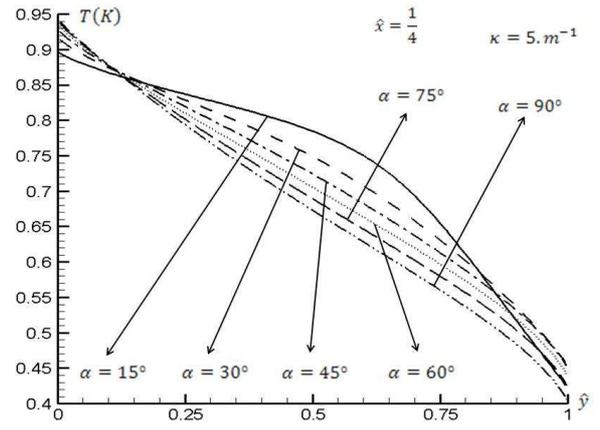
(a2)

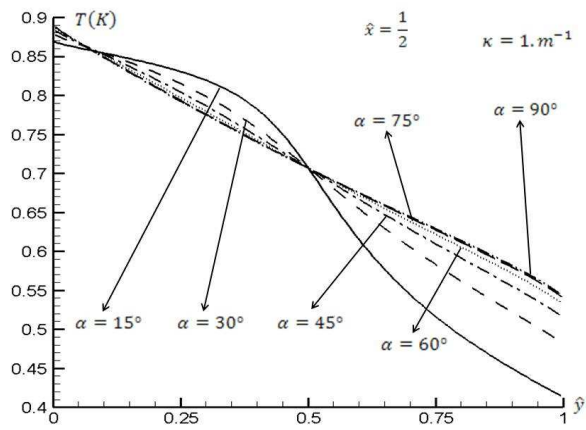
(b1)

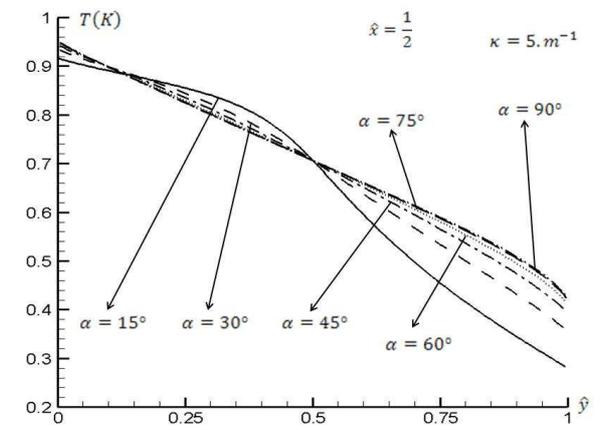
(b2)

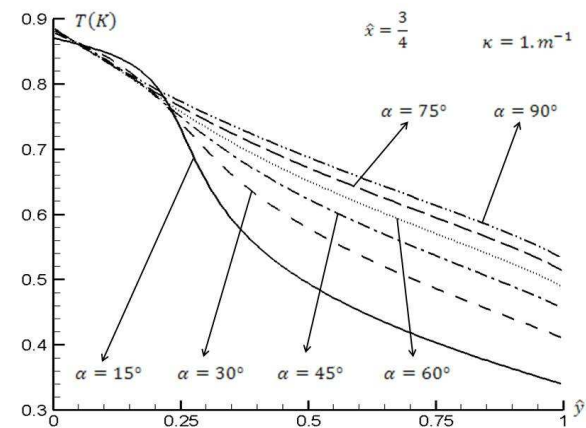
(c1)

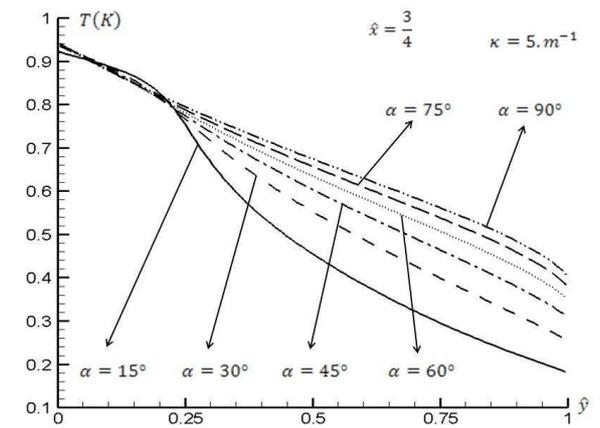
(c2)

Figure 4: Temperatures on the lines $\hat{x} = \left\{\frac{1}{4}(a), \frac{1}{2}(b), \frac{3}{4}(c)\right\}$ for different angles $\alpha = \{15°, 30°, 45°, 60°, 75°, 90°\}$ for two absorption coefficients $\kappa = \{1.(1), 5.(2)\} m^{-1}$

It is noticeable that the temperature on the line $\hat{x} = \frac{3}{4}$ in the neighbourhood of the top boundary is significantly lower for cavities with small angles $\alpha$ than for square cavities. On the medium line of the cavity (i.e. $\hat{x} = \frac{1}{2}$), except for $\alpha = 15°$, the temperatures discrepancy between the different cases is not very large. It is possible to show, at least for the optically thin approximation, that for angles lower than 30°, the temperatures and radiative flux profiles are different from profiles obtained for higher angles, which can explain the exception at $\alpha = 15°$.

When we use a numerical method to solve a transport-diffusion equation in a domain having complex geometry, we have to deal with skewed non-orthogonal meshes or grids (which is the case in this work). In this situation, it is well known, when radiation does not occur, that the precision of numerical schemes degrades. As the skewness is more pronounced for small angles we shall therefore present the temperature and flux fields for only three angles $= \{15°, 30°, 45°\}$ and for the square cavity. Temperature fields are depicted in Figure 5 for the two absorption coefficients $\kappa = \{1.(1), 5.(2)\}m^{-1}$, when the non-zero temperature of the southern boundary is set to $T_S = 1000K$. They have similar trends whatever the edge angle is, except in the neighbourhood of the top surface for very small angles. It is seen from these figures that isotherms are equally spaced between the maximum temperature of the hot wall and the minimum temperature of the cold walls. It is worth to note the absence of any numerical oscillation.

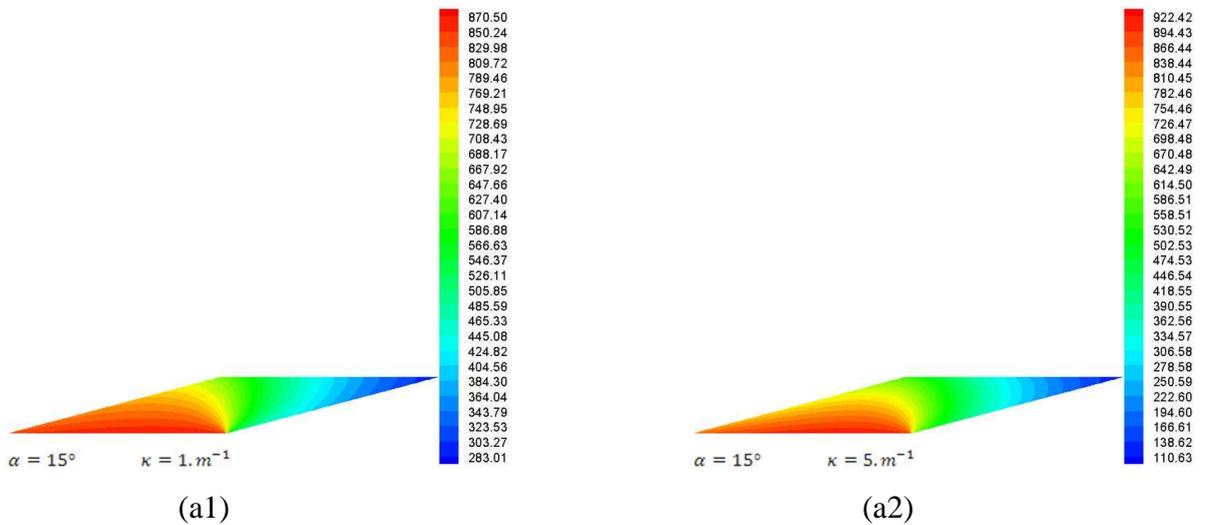

(a1)                                                    (a2)

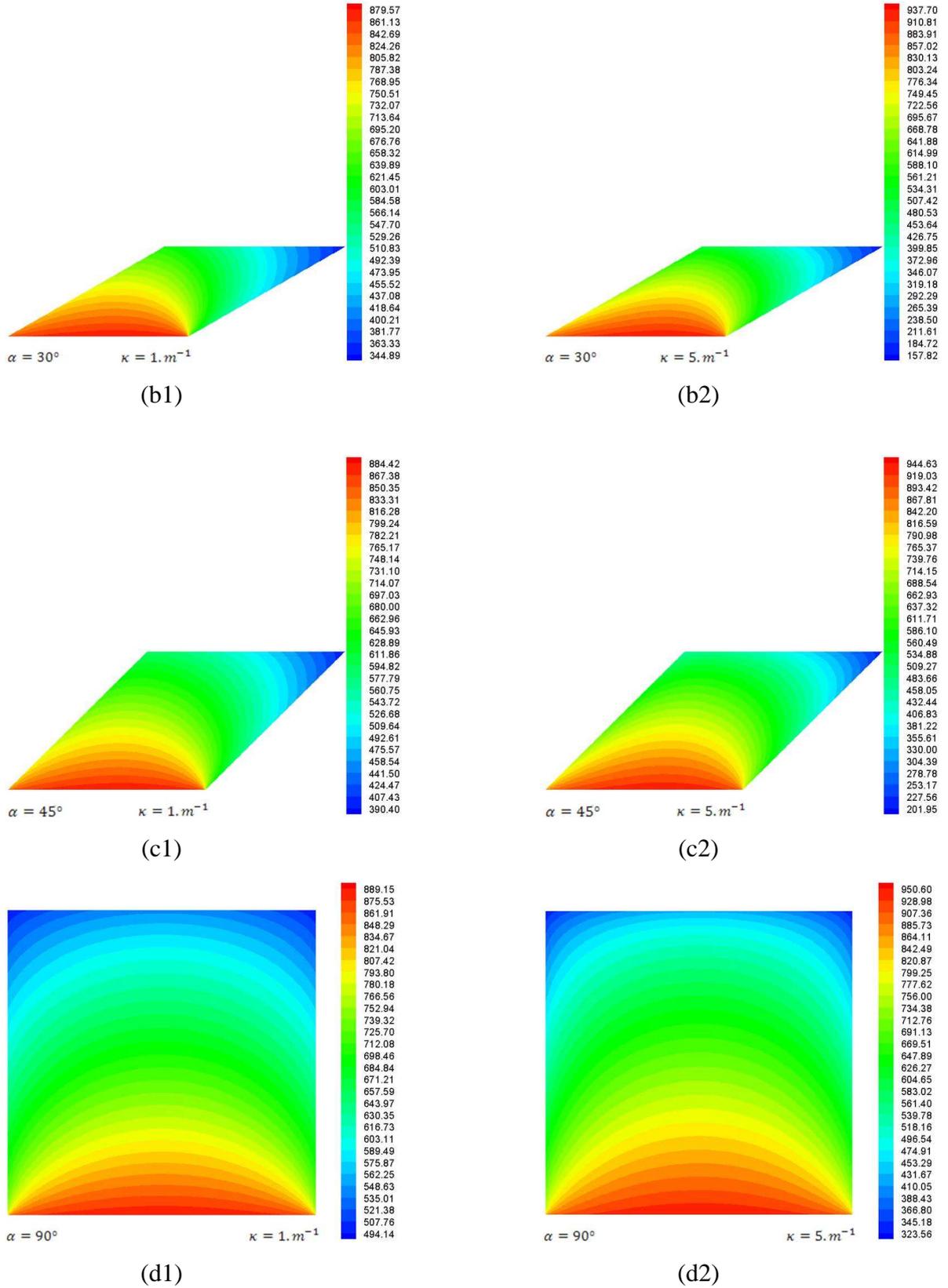

Figure 5: Temperature fields for different angles $\alpha = \{15°(a), 30°(b), 45°(c), 90°(d)\}$ for two absorption coefficients $\kappa = \{1.(1), 5.(2)\}m^{-1}$

Despite their physical significance, it appears that numerical results focusing on radiative fluxes have not been sufficiently reported. We therefore give in figures 6 and 7 the two flux components of the reduced radiative flux, defined as $\vec{q_r^*} = \frac{\vec{q_r}}{\sigma T_S^4}$, corresponding to the same parameters as those of figure 5. Due to the boundary conditions on the surfaces, the first component presents positive and negative values. In the case of the square cavity, the medium line $\hat{x} = \frac{1}{2}$ is the line of zero value for this component and the numerical values of the *x*-flux for the eastern part of the cavity are exactly opposite to the ones located in the western part relatively to the symmetry line. For angles lower than 90°, the 0 constant-line is no longer a straight line and its location depends also on the absorption coefficient, contrarily to the square cavity. The positive and negative values of the *x*-component are not symmetric in the general case of a parallelogrammic cavity.

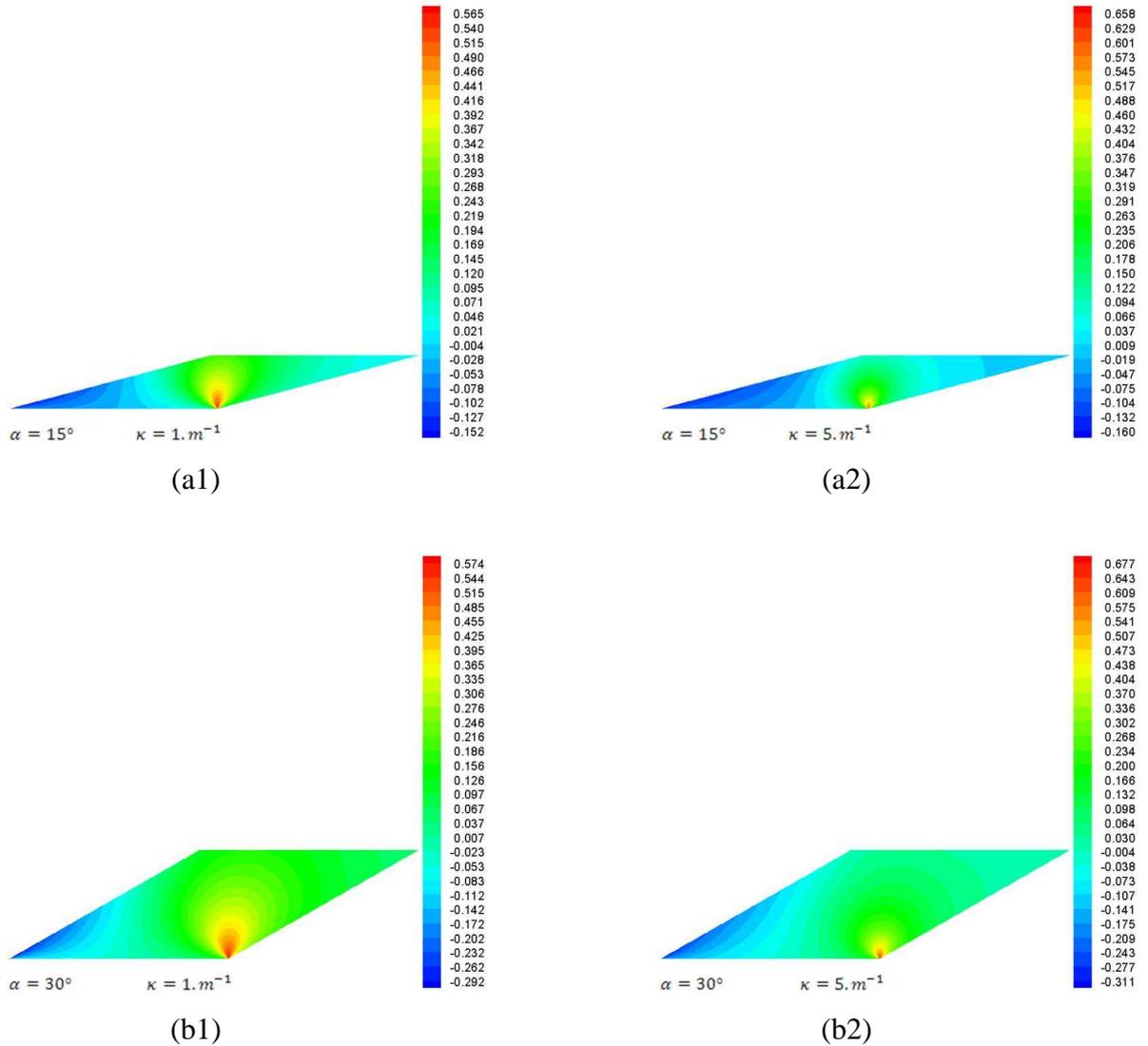

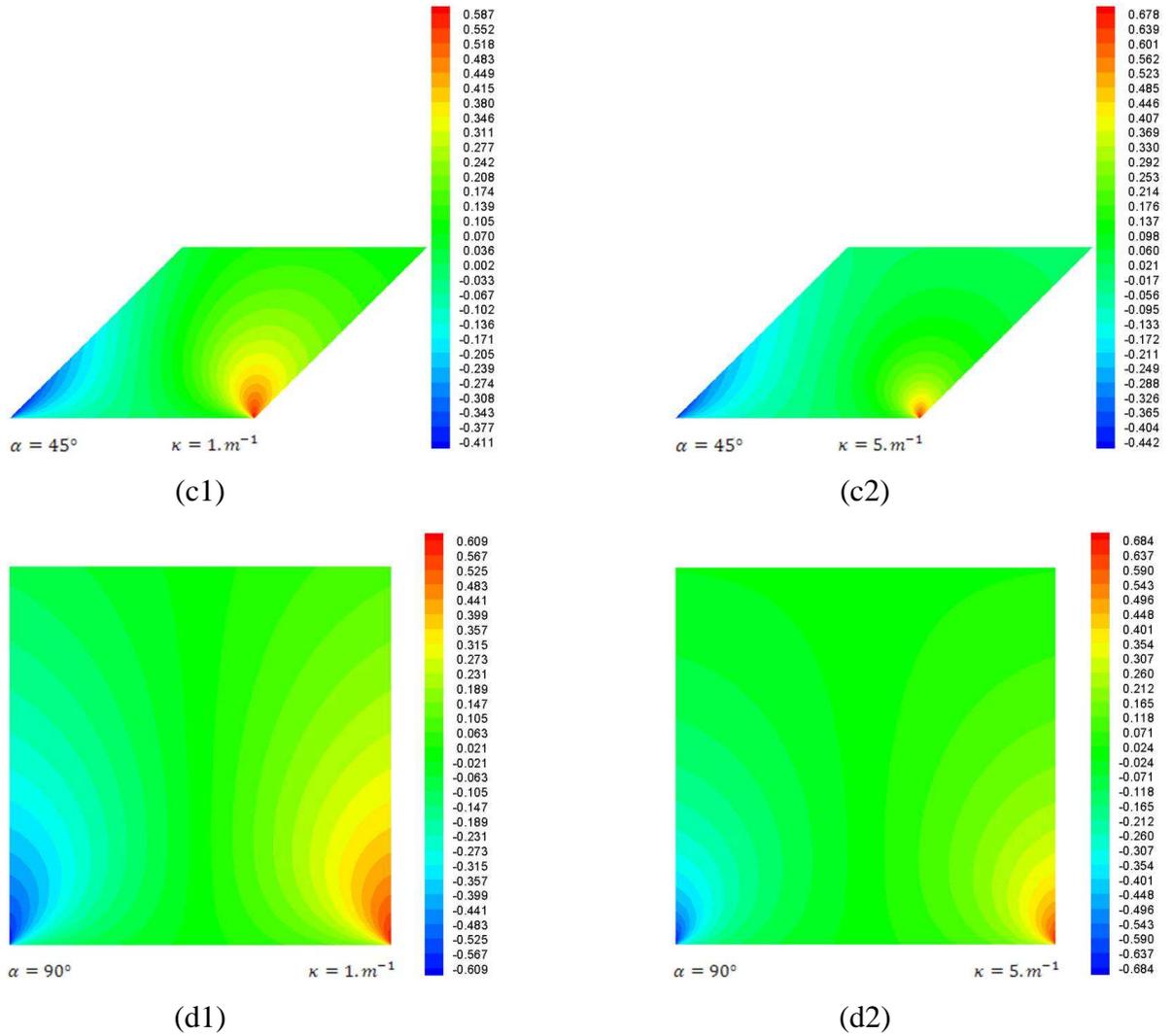

Figure 6: *x*-component of radiative flux for different angles $\alpha = \{15°(a), 30°(b), 45°(c), 90°(d)\}$ and two absorption coefficients $\kappa = \{1.(1), 5.(2)\} m^{-1}$

The *y*-component of the radiative flux field is presented in figure 7. It is also very different for small angles from the one obtained in a square cavity.

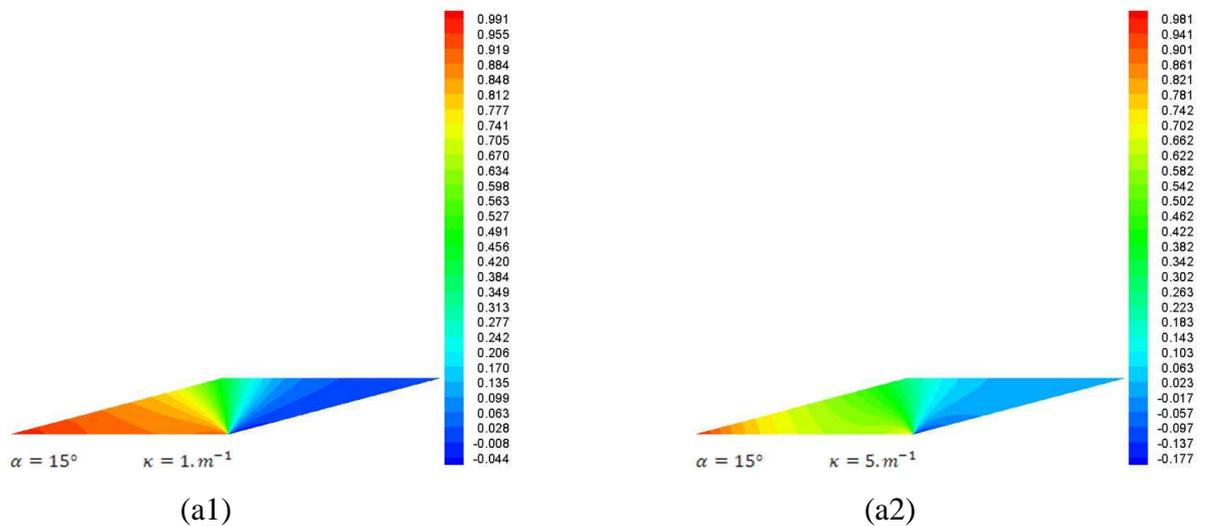

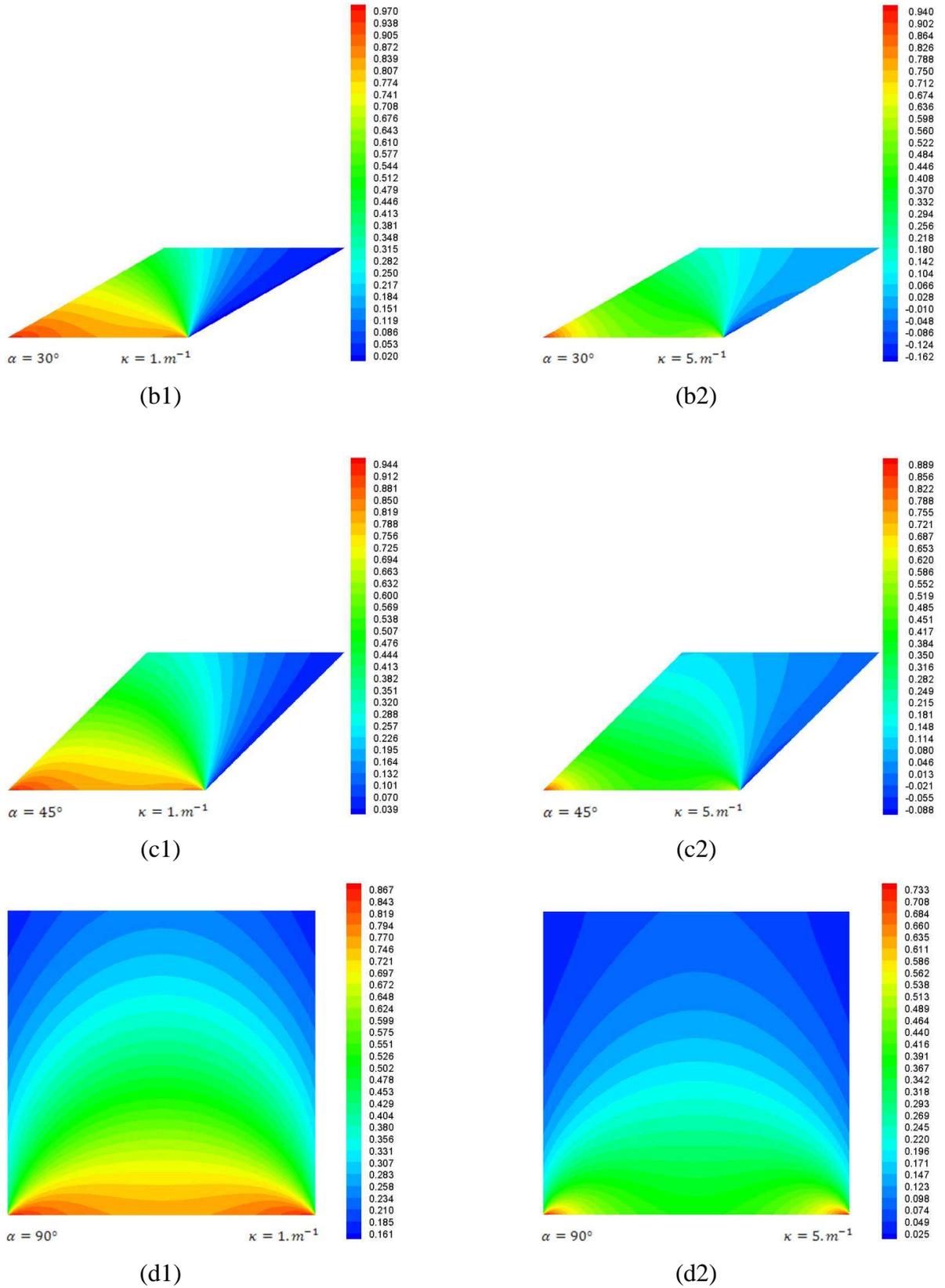

Figure 7: *y*-component of radiative flux for different angles $\alpha = \{15°(a), 30°(b), 45°(c), 90°(d)\}$ and two absorption coefficients $\kappa = \{1.(1), 5.(2)\}m^{-1}$

## III.2 Prescribed temperature field

We now turn to the case of a medium with a prescribed temperature field given by the analytical expression:

$$T(\hat{x}, \hat{y}) = \frac{4T_S}{\pi} \sum_{n=1}^{+\infty} \frac{\sin[(2n-1)\pi\hat{x}] \sinh[(2n-1)\pi(1-\hat{y})]}{(2n-1) \sinh[(2n-1)\pi]}$$

where $T_S$ is the temperature of the southern surface of the cavity. This is an extension to the parallelogram cavity of the steady heat conduction field in a square cavity. The incident radiation $G$ is then calculated and put in an adimensional form by introducing the apparent internal temperature given by the expression:

$T_a(\hat{x}, \hat{y}) = \frac{1}{T_S} \sqrt[4]{\frac{G(\hat{x},\hat{y})}{4\sigma}}$. These two fields are presented on Figure 8 for an angle a=45° and κ=1m$^{-1}$ when the non-zero temperature of the southern boundary is set to $T_S = 1000K$.

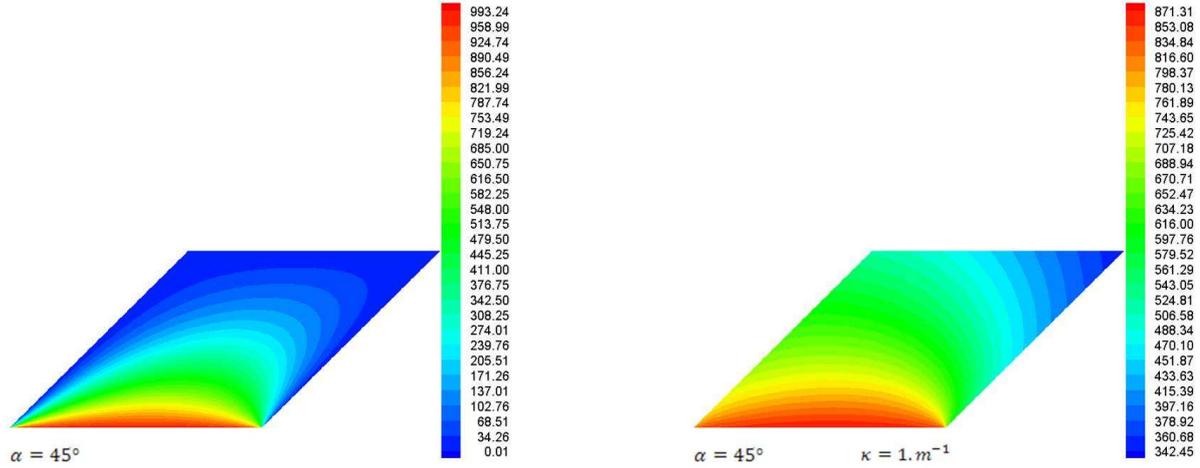

Figure 8: Prescribed temperature field and apparent temperature field for different angles $\alpha = 45°$ and $\kappa = 1m^{-1}$

We give in Tables 2a and 2b discrete values of the non-dimensional incident radiation on the medium line of the cavity $\hat{x} = \frac{1}{2}$ at regularly spaced ordinates positions $\hat{y}$ for six values of the edge angle $\alpha$ and for two absorption coefficients identical to the ones used in the previous study at radiative equilibrium.

| $\hat{y}$ | $\alpha = 15°$ | $\alpha = 30°$ | $\alpha = 45°$ | $\alpha = 60°$ | $\alpha = 75°$ | $\alpha = 90°$ |
|---|---|---|---|---|---|---|
| 0.05 | 0.8473 | 0.8500 | 0.8514 | 0.8522 | 0.8526 | 0.8527 |
| 0.10 | 0.8348 | 0.8293 | 0.8247 | 0.8213 | 0.8193 | 0.8186 |
| 0.15 | 0.8237 | 0.8107 | 0.8008 | 0.7939 | 0.7898 | 0.7885 |
| 0.20 | 0.8107 | 0.7891 | 0.7735 | 0.7630 | 0.7571 | 0.7551 |
| 0.25 | 0.7987 | 0.7696 | 0.7496 | 0.7368 | 0.7297 | 0.7275 |
| 0.30 | 0.7831 | 0.7459 | 0.7224 | 0.7081 | 0.7004 | 0.6979 |
| 0.35 | 0.7663 | 0.7235 | 0.6987 | 0.6841 | 0.6763 | 0.6739 |
| 0.40 | 0.7404 | 0.6954 | 0.6717 | 0.6580 | 0.6507 | 0.6483 |
| 0.45 | 0.7079 | 0.6689 | 0.6485 | 0.6363 | 0.6297 | 0.6276 |
| 0.50 | 0.6621 | 0.6394 | 0.6243 | 0.6144 | 0.6087 | 0.6068 |
| 0.55 | 0.6113 | 0.6099 | 0.6010 | 0.5935 | 0.5888 | 0.5873 |
| 0.60 | 0.5676 | 0.5837 | 0.5803 | 0.5752 | 0.5715 | 0.5702 |
| 0.65 | 0.5254 | 0.5557 | 0.5581 | 0.5555 | 0.5529 | 0.5519 |
| 0.70 | 0.4947 | 0.5331 | 0.5397 | 0.5391 | 0.5374 | 0.5367 |
| 0.75 | 0.4653 | 0.5095 | 0.5201 | 0.5214 | 0.5208 | 0.5203 |
| 0.80 | 0.4435 | 0.4908 | 0.5040 | 0.5069 | 0.5070 | 0.5068 |
| 0.85 | 0.4218 | 0.4711 | 0.4868 | 0.4912 | 0.4920 | 0.4920 |
| 0.90 | 0.4050 | 0.4553 | 0.4726 | 0.4781 | 0.4795 | 0.4797 |
| 0.95 | 0.3880 | 0.4387 | 0.4574 | 0.4639 | 0.4659 | 0.4663 |

Table 2a: Equivalent temperature $T_a$ on the line $\hat{x} = \frac{1}{2}$ for different angles $\alpha$ at $\kappa = 1.\,m^{-1}$

| $\hat{y}$ | $\alpha = 15°$ | $\alpha = 30°$ | $\alpha = 45°$ | $\alpha = 60°$ | $\alpha = 75°$ | $\alpha = 90°$ |
|---|---|---|---|---|---|---|
| 0.05 | 0.8679 | 0.8782 | 0.8838 | 0.8869 | 0.8886 | 0.8890 |
| 0.10 | 0.8345 | 0.8293 | 0.8254 | 0.8227 | 0.8212 | 0.8207 |
| 0.15 | 0.8048 | 0.7851 | 0.7723 | 0.7643 | 0.7600 | 0.7586 |
| 0.20 | 0.7714 | 0.7350 | 0.7123 | 0.6985 | 0.6912 | 0.6889 |
| 0.25 | 0.7427 | 0.6922 | 0.6614 | 0.6432 | 0.6337 | 0.6307 |
| 0.30 | 0.7102 | 0.6445 | 0.6060 | 0.5839 | 0.5724 | 0.5688 |
| 0.35 | 0.6807 | 0.6037 | 0.5604 | 0.5359 | 0.5233 | 0.5193 |
| 0.40 | 0.6425 | 0.5573 | 0.5114 | 0.4855 | 0.4721 | 0.4679 |
| 0.45 | 0.6006 | 0.5166 | 0.4713 | 0.4453 | 0.4317 | 0.4275 |
| 0.50 | 0.5443 | 0.4736 | 0.4315 | 0.4063 | 0.3928 | 0.3885 |
| 0.55 | 0.4815 | 0.4318 | 0.3945 | 0.3707 | 0.3577 | 0.3535 |
| 0.60 | 0.4262 | 0.3951 | 0.3629 | 0.3409 | 0.3284 | 0.3244 |
| 0.65 | 0.3717 | 0.3565 | 0.3301 | 0.3101 | 0.2984 | 0.2946 |
| 0.70 | 0.3313 | 0.3256 | 0.3037 | 0.2856 | 0.2747 | 0.2711 |
| 0.75 | 0.2926 | 0.2940 | 0.2765 | 0.2604 | 0.2504 | 0.2471 |
| 0.80 | 0.2639 | 0.2693 | 0.2550 | 0.2406 | 0.2314 | 0.2282 |
| 0.85 | 0.2357 | 0.2440 | 0.2326 | 0.2201 | 0.2117 | 0.2088 |
| 0.90 | 0.2143 | 0.2241 | 0.2149 | 0.2037 | 0.1960 | 0.1934 |
| 0.95 | 0.1930 | 0.2037 | 0.1966 | 0.1868 | 0.1798 | 0.1774 |

Table 2b: Equivalent temperature $T_a$ on the line $\hat{x} = \frac{1}{2}$ for different angles $\alpha$ at $\kappa = 5.\,m^{-1}$

We finally present on figure 9 the apparent temperatures profiles for different angles $\alpha$ and on the two lines $\hat{x} = \frac{1}{4}$ and $\hat{x} = \frac{3}{4}$ for $\kappa = 1.\,m^{-1}$. These profiles are not symmetrical as it is the case in the square cavity.

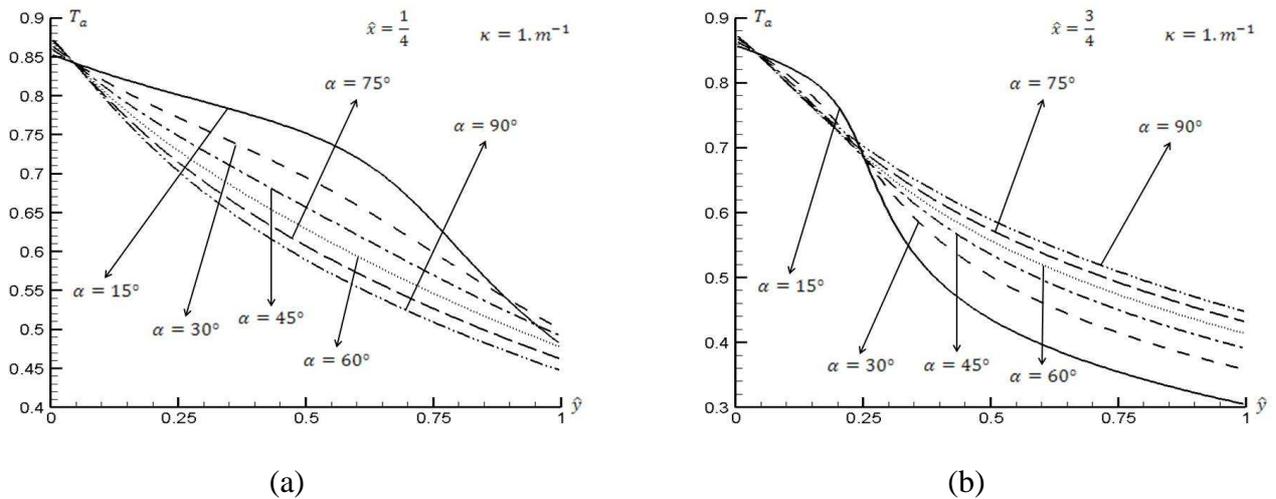

(a)  (b)

Figure 9: Reduced apparent temperatures $T_a$ on the lines $\hat{x} = \left\{\frac{1}{4}(a), \frac{3}{4}(b)\right\}$ for different angles $\alpha = \{15°, 30°, 45°, 60°, 75°, 90°\}$ and for $\kappa = 1\,m^{-1}$

Since the global energy equation is not verified, the apparent temperature fields are significantly different from the prescribed one, especially for low absorption coefficients. If one compares figures 5 and 8, the apparent temperature field and the radiative equilibrium temperature field are qualitatively and quantitatively quite similar in the neighbourhood of the southern surface but strongly different in the other locations of the cavity.

**IV – CONCLUDING REMARKS**

In this paper we have presented an efficient numerical technique based on analytical developments to calculate the complete internal radiative field inside a grey semi-transparent medium enclosed in a parallelogram shaped cavity bounded by black surfaces. The method yields to smooth temperature and flux fields without any oscillatory behaviour. The obtained numerical results are extremely reliable for small and moderate optical depths. Besides, these results can be used as benchmark solutions to assess the validity of numerical methods especially for small angles. The determination of the temperature and radiative flux profiles for cavities with diffusely reflecting surfaces is in progress.

**Annex 1**

The different boundary terms write with the help of the two successive variable changes $\xi = \alpha - \varphi$ and $\omega = \frac{\pi}{2} - \xi$:

$$\int_{\varphi=0}^{\varphi_E^+} Ki_2\left[\frac{\kappa(H_x - \hat{x})\sin\alpha}{\sin(\alpha - \varphi)}\right] d\varphi = Bis_2[\kappa(H_x - \hat{x})\sin\alpha, \psi_E^+] - Bis_2\left[\kappa(H_x - \hat{x})\sin\alpha, \frac{\pi}{2} - \alpha\right]$$

where $\psi_E^+ = \tan^{-1}\left[\frac{H_y - \hat{y} + (H_x - \hat{x})\cos\alpha}{(H_x - \hat{x})\sin\alpha}\right]$, and:

$$\int_{\varphi=0}^{\varphi_E^+} Ki_3\left[\frac{\kappa(H_x - \hat{x})\sin\alpha}{\sin(\alpha - \varphi)}\right] \cos\varphi\, d\varphi$$
$$= \sin\alpha \left\{Bis_3[\kappa(H_x - \hat{x})\sin\alpha, \psi_E^+] - Bis_3\left[\kappa(H_x - \hat{x})\sin\alpha, \frac{\pi}{2} - \alpha\right]\right\}$$
$$+ \cos\alpha \left\{Cis_3[\kappa(H_x - \hat{x})\sin\alpha, \psi_E^+] - Cis_3\left[\kappa(H_x - \hat{x})\sin\alpha, \frac{\pi}{2} - \alpha\right]\right\}$$

$$\int_{\varphi=0}^{\varphi_E^+} Ki_3\left[\frac{\kappa(H_x - \hat{x})\sin\alpha}{\sin(\alpha - \varphi)}\right] \sin\varphi\, d\varphi$$
$$= \sin\alpha \left\{Cis_3[\kappa(H_x - \hat{x})\sin\alpha, \psi_E^+] - Cis_3\left[\kappa(H_x - \hat{x})\sin\alpha, \frac{\pi}{2} - \alpha\right]\right\}$$
$$- \cos\alpha \left\{Bis_3[\kappa(H_x - \hat{x})\sin\alpha, \psi_E^+] - Bis_3\left[\kappa(H_x - \hat{x})\sin\alpha, \frac{\pi}{2} - \alpha\right]\right\}$$

Similarly to what precedes we have:

$$\int_{\varphi=0}^{\varphi_{\overline{E}}} Ki_2\left[\frac{\kappa(H_x-\hat{x})\sin\alpha}{\sin(\alpha+\varphi)}\right]d\varphi = Bis_2\left[\kappa(H_x-\hat{x})\sin\alpha,\frac{\pi}{2}-\alpha\right] + \delta_E Bis_2\{\kappa(H_x-\hat{x})\sin\alpha,\omega_E\}$$

$$\int_{\varphi=0}^{\varphi_{\overline{E}}} Ki_3\left[\frac{\kappa(H_x-\hat{x})\sin\alpha}{\sin(\alpha+\varphi)}\right]\cos\varphi\, d\varphi$$
$$= \sin\alpha\, \langle Bis_3\left[\kappa(H_x-\hat{x})\sin\alpha,\frac{\pi}{2}-\alpha\right] + \delta_E Bis_3\{\kappa(H_x-\hat{x})\sin\alpha,\omega_E\}\rangle$$
$$+ \cos\alpha\, \langle Cis_3\left[\kappa(H_x-\hat{x})\sin\alpha,\frac{\pi}{2}-\alpha\right] - Cis_3\{\kappa(H_x-\hat{x})\sin\alpha,\omega_E\}\rangle$$

$$\int_{\varphi=0}^{\varphi_{\overline{E}}} Ki_3\left[\frac{\kappa(H_x-\hat{x})\sin\alpha}{\sin(\alpha+\varphi)}\right]\sin\varphi\, d\varphi$$
$$= \sin\alpha\, \langle Cis_3\left[\kappa(H_x-\hat{x})\sin\alpha,\frac{\pi}{2}-\alpha\right] - Cis_3\{\kappa(H_x-\hat{x})\sin\alpha,\omega_E\}\rangle$$
$$- \cos\alpha\, \langle Bis_3\left[\kappa(H_x-\hat{x})\sin\alpha,\frac{\pi}{2}-\alpha\right] + \delta_E Bis_3\{\kappa(H_x-\hat{x})\sin\alpha,\omega_E\}\rangle$$

Where: $\omega_E = \begin{cases} \tan^{-1}\left[\frac{\hat{y}-(H_x-\hat{x})\cos\alpha}{(H_x-\hat{x})\sin\alpha}\right] \\ \tan^{-1}\left[\frac{(H_x-\hat{x})\cos\alpha-\hat{y}}{(H_x-\hat{x})\sin\alpha}\right] \end{cases}$ and $\delta_E = \begin{cases} 1 & \text{if } \hat{y} \geq (H_x-\hat{x})\cos\alpha \\ -1 & \text{if } \hat{y} < (H_x-\hat{x})\cos\alpha \end{cases}$

$Bis_2$, $Bis_3$ and $Cis_3$ are the incomplete integrated Bickley-Naylor functions [8].

## Annex 2

If we consider as a matter of example the incident radiation from the northern part of the eastern surface, the first integral in Eq. (13) writes:

$$\int_{\varphi=0}^{\varphi_1}\int_{\hat{u}=\hat{x}_i}^{H_x} T^4\left[\hat{u},\hat{y}_j + \frac{(\hat{u}-\hat{x}_i)\sin\varphi}{\sin(\alpha-\varphi)}\right] Ki_1\left[\frac{\kappa(\hat{u}-\hat{x}_i)\sin\alpha}{\sin(\alpha-\varphi)}\right] d\hat{u}\, \frac{d\varphi}{\sin(\alpha-\varphi)}$$
$$= (H_x-\hat{x}_i)\varphi_1$$
$$\times \sum_{l=1}^{N_\varphi}\sum_{m=1}^{N_u} \frac{\omega_l\omega_m}{\sin(\alpha-\varphi_l)} T^4\left[\hat{x}_i + (H_x-\hat{x}_i)\beta_m, \hat{y}_j\right.$$
$$\left. + \frac{(H_x-\hat{x}_i)\beta_m \sin\varphi_l}{\sin(\alpha-\varphi_l)}\right] Ki_1\left[\frac{\kappa(H_x-\hat{x}_i)\beta_m \sin\alpha}{\sin(\alpha-\varphi_l)}\right]$$

where $\varphi_1 = \tan^{-1}\left[\frac{(H_y-\hat{y}_j)\sin\alpha}{H_x-\hat{x}_i+(H_y-\hat{y}_j)\cos\alpha}\right]$ and $\varphi_l = \varphi_1\beta_l$, the $\beta_l$ (respectively $\beta_m$) being the abscissa and the $\omega_l, \omega_m$ the weights of the numerical quadratures of $N_\varphi$ and $N_u$ points used to compute the double angular and spatial integral. In the previous expression, $\hat{x}_i$ and $\hat{y}_j$ are the coordinates of the centre of the quadrilateral cell $(i,j)$ with $\hat{x}_1 = 0$, $\hat{x}_i = \frac{i-\frac{3}{2}}{N_x-2}H_x$ for $2 \leq i \leq N_x-1$, $\hat{x}_{N_x} = H_x$ and $\hat{y}_1 = 0$, $\hat{y}_j = \frac{j-\frac{3}{2}}{N_y-2}H_y$ for $2 \leq j \leq N_y-1$, $\hat{y}_{N_y} = H_y$. Each point of a cell $(i,j)$ being at the same temperature than the centre $(\hat{x}_i, \hat{y}_j)$, there exist two integers $p$ and $q$ depending on $i, j, l$ and $m$ such that:

$$\hat{x}_p - \frac{\Delta \hat{x}}{2} \leq \hat{x}_i + (H_x - \hat{x}_i)\beta_m < \hat{x}_p + \frac{\Delta \hat{x}}{2}$$

$$\hat{y}_q - \frac{\Delta \hat{y}}{2} \leq \hat{y}_j + \frac{(H_x - \hat{x}_i)\beta_m \sin \varphi_l}{\sin(\alpha - \varphi_l)} < \hat{y}_q + \frac{\Delta \hat{y}}{2}$$

and defined by:

$$p = \left[i + \frac{1}{2} + \beta_m \left(N_x - i - \frac{1}{2}\right)\right]$$

$$q = \left[j + \frac{1}{2} + \frac{\beta_m \left(N_x - i - \frac{1}{2}\right) \Delta \hat{x} \sin \varphi_l}{\Delta \hat{y} \sin(\alpha - \varphi_l)}\right]$$

where $[x]$ denotes the integer part of the real number $x$. Then the discrete expression of the integral rewrites:

$$\int_{\varphi=0}^{\varphi_1} \int_{\hat{u}=\hat{x}_i}^{H_x} T^4 \left[\hat{u}, \hat{y}_j + \frac{(\hat{u} - \hat{x}_i)\sin \varphi}{\sin(\alpha - \varphi)}\right] Ki_1 \left[\frac{\kappa(\hat{u} - \hat{x}_i)\sin \alpha}{\sin(\alpha - \varphi)}\right] d\hat{u} \frac{d\varphi}{\sin(\alpha - \varphi)}$$

$$= (H_x - \hat{x}_i)\varphi_1 \sum_{l=1}^{N_\varphi} \sum_{m=1}^{N_u} \frac{\omega_l \omega_m}{\sin(\alpha - \varphi_l)} T_{pq}^4 Ki_1 \left[\frac{\kappa(H_x - \hat{x}_i)\beta_m \sin \alpha}{\sin(\alpha - \varphi_l)}\right]$$

A similar treatment is performed for all the integrals of the incident radiation and radiative flux.